\documentclass[12pt]{iopart}
\usepackage{graphicx}
\usepackage{iopams}
\usepackage{bbm}

\newcommand{\op}[2]{\left| #1 \right\rangle \left\langle #2 \right|}
\newcommand{\ket}[1]{\left| #1 \right\rangle}

\newcommand{\EV}[1]{\left\langle #1 \right\rangle}
\newcommand{\dA}{\delta\! A}
\newcommand{\dt}{\frac{\rmd}{\rmd t}}

\begin{document}
\title{Cavity Assisted Nondestructive Laser Cooling of Atomic Qubits}

\author{A. Griessner$^\dag$, D.~Jaksch$^{\ddag}$ and P.~Zoller$^{\dag,\P}$}

\address{$\dag$ Institute for Theoretical Physics, University
of Innsbruck, A--6020 Innsbruck, Austria.}

\address{$\ddag$ Clarendon Laboratory, Department of Physics,
University of Oxford, Oxford, United Kingdom}

\address{$\P$ Institute for Quantum Optics and Quantum
Information of the Austrian Academy of Sciences, A-6020
Innsbruck, Austria.}

\begin{abstract}
We analyze two configurations for laser cooling of neutral atoms
whose internal states store qubits. The atoms are trapped
in an optical lattice which is placed inside a cavity. We
show that the coupling of the atoms to the damped cavity mode can
provide a mechanism which leads to cooling of the motion without
destroying the quantum information.
\end{abstract}

\maketitle

\section{Introduction}

Quantum optical schemes for quantum information processing employ
long lived internal (spin or hyperfine)  states of atoms or ions
to store the qubit \cite{PZReview}. In many of the proposals for
scalable quantum computing these qubits are stored in a storage
area, and must be moved together to perform e.g.~a two-qubit gate
operation entangling the atoms \cite{WinelandScale,PZMicro}.
Moving atoms will  typically cause heating, and thus the question
arises to what extent the motion of the qubit atoms can be cooled
{\em nondestructively}, i.e.~without affecting the coherence of
the qubit. In the case of quantum computing with trapped ions this
is achieved by sympathetic cooling \cite{WinelandSymp}: the
``qubit ion'' is coupled via the Coulomb force to a second
``cooling ion'' (of a different species), which is laser cooled
\cite{LaserBlatt,JosaMetcalf}. This cooling is nondestructive
because the Coulomb force is identical for the two qubit states
$|0\rangle$, $|1\rangle$. Quantum computing with neutral atoms can
be performed by storing atoms in optical lattices generated by
counterpropagating laser beams.  By moving the atoms in a spin
dependent optical lattice \cite{ColdColl,BlochMove} one can induce
controlled collisions between the qubits leading to multi particle
entanglement \cite{ColdColl,BlochEnt,BrennenQLneutral}. One
possibility for nondestructive cooling in an optical lattice is
sympathetic collisional cooling by immersion of the qubit atoms
stored in the lattice in an atomic Bose Einstein condensate
\cite{Andrew}. Of course, laser cooling of qubit atoms stored in
the wells of an optical lattice potential cannot be directly
applied, as the optical pumping associated with the laser cooling
process destroys the coherence. Instead we will investigate below
a scenario of nondestructive laser cooling of atomic qubits in
moving optical lattice potentials which is achieved by coupling to
a (low-$Q$) optical cavity. These results are also of interest for
implementations of quantum computation and quantum communication
with optical cavities
\cite{Pellizzari,Parkins,Mabuchi,Grangier,Hemmerich}.

Cavity assisted laser cooling is a well-studied process
\cite{RitschReview,Vuletic,Rempe,Kimble,HZHighQ,HZOL,HemmerichCoolRing}, both
theoretically and experimentally. In the present paper we will
elaborate on two distinguishing features which make cavity
assisted laser cooling of interest for quantum computing with
atoms in optical lattices. We will study a scenario where a set of
qubit atoms is stored in individual wells of an optical lattice
potential. By moving the lattice we will heat the atom motion.
However, to the extent the optical lattice potentials seen by the
two qubit states are identical, this transport will not destroy
the coherence of the internal atomic states. Cavity assisted
recooling of the motional degrees works by converting phonons to
cavity photons in a coherent laser assisted process. These cavity
photons dissipate through the cavity mirror. With an appropriate
choice of internal qubit states, this cooling process does not
distinguish between the two qubits, and is thus nondestructive. We
also note that moving $N$ qubits in an accelerating lattice will
excite the collective center of mass mode of these atoms. This is
precisely the mode which is cooled most efficiently by our cavity
cooling scheme. The scheme we investigate is, in principle,
capable to achieve ground state cooling in the optical wells,
although in practice nondestructive cooling to the Lamb-Dicke
regime, where the atom is localized in a region much smaller than
the optical wave length, may be sufficient.

The paper is organized as follows. In Sec.~\ref{model} we present
two different specific setups and we calculate the
equations of motion in both cases for a single
qubit and for the center of mass motion of N qubits. In
Sec.~\ref{Cool} we derive analytical expressions for the steady
state temperature $T_f$ and the cooling time $\tau$ of the atoms
and compare them with numerical calculations. Finally, in
Sec.~\ref{imperfect} we discuss possible imperfections of the
system.

\section{The Model}\label{model}

In this section we introduce two different setups for cooling the
motion of neutral atoms without destroying the quantum information
stored in two long lived degenerate internal ground states
$\ket{g_1} \equiv \ket{0}$ and $\ket{g_2} \equiv \ket{1}$.  In the
first setup the atoms are trapped and cooled by the field of a
ring cavity while in the second case trapping is achieved by an
optical lattice and the cavity is solely used for cooling the
atoms. To be specific, we will choose the two qubit states to be
two Zeeman states, e.g.~$|J=1/2,M=\pm 1/2\rangle$, and we will
assume identical (symmetric) coupling of these states to the laser
and the cavity mode to achieve cooling which does not distinguish
between the two qubit states. Our goal is to derive cooling
equations first for a single qubit, and then for the center of
mass motion of $N$ qubits.

\subsection{Optical Lattice in a Ring Cavity}
\label{model_ring}

We consider neutral atoms in a ring cavity as shown in
Fig.~\ref{fig:RingFig}a). Two coherent fields $\beta_R$ and
$\beta_L$ drive two counter-propagating running waves adding up to
a standing wave inside the cavity. We first consider a single
qubit in this ring cavity and then extend the system to $N$ qubits.

\subsubsection{Single Qubit}\label{singleQ}
\begin{figure}[htp]
    \begin{center}
    \includegraphics{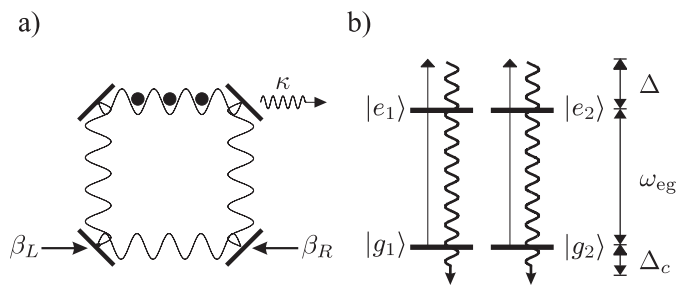}
    \caption{\label{fig:RingFig} a) Experimental setup as described
     in the text. b) Schematic
    picture of the internal level structure, the involved frequencies
    and detunings. The ground state $\ket{g_1}$ $(\ket{g_2})$ is identified
    with the qubit state $\ket{0}$ $(\ket{1})$. As a concrete example one can
    chose the fine structure states (e.g.~of Na or Rb) as shown in
    Fig.~\ref{fig:Alkali}.}
    \end{center}
\end{figure}

In a frame rotating with the laser frequency $\omega_L$, the Hamiltonian describing
the cavity fields and one atom is given by
\begin{equation}\label{Hsum}
   H=H_{\rm A} +H_{\rm C}+H_{\rm AC}+H_{\rm F}+H_{\rm CF}+H_{\rm bath}+H_{\rm
    sys-bath}
\end{equation}
with ($\hbar=1$)
\begin{eqnarray}\label{Hexplain}
    &H_{\rm A}
    =\frac{p^2}{2m}+(-\Delta)\left(\op{e_1}{e_1}+\op{e_2}{e_2}\right),\\
    &H_{\rm C}=(-\Delta_c)(a_R^{\dagger}a_R+a_L^{\dagger}a_L),\nonumber\\
     &H_{\rm AC}=g\left[\left(a_R\rme^{\rmi(kx-\omega_L
     t)}+a_L\rme^{-\rmi(kx+\omega_L t)}\right)\left(\sigma_1^++\sigma_2^+\right)+{\rm
     h.c.}\right],\nonumber\\
     &H_{\rm F}=\sum_{j=R,L}\int {\rm d}\omega \omega b_{\omega,j}^\dagger
     b_{\omega,j},\nonumber\\
     &H_{\rm CF}=\rmi\sqrt{\kappa}\sum_{j=R,L}\int{\rm d}\omega \left(a_j^\dagger
     b_{\omega,j}\rme^{\rmi\omega_L t}-{\rm h.c.}\right).\nonumber
\end{eqnarray}
Here the detuning of the laser from the atomic transition
frequency $\omega_{\rm eg}$ is given by
$\Delta=\omega_L-\omega_{\rm eg}$ and the detuning from the cavity
frequency $\omega_c$ by $\Delta_c=\omega_L-\omega_c$. Furthermore,
$H_{\rm A}$ gives the kinetic and the internal energy of the atom,
respectively, where $p$ is the momentum operator and $m$ is the
mass of the atom. The two long lived degenerate ground states
carrying the qubit are $\ket{g_1} \equiv \ket{0}$ and $\ket{g_2}
\equiv \ket{1}$, as sketched in Fig.~\ref{fig:RingFig} and we
assume the atom to have two degenerate excited states $\ket{e_1}$
and $\ket{e_2}$. The Hamiltonian $H_{\rm C}$ describes a ring
cavity with a clockwise and counter-clockwise propagating mode,
and bosonic destruction (creation) operators $a_{R(L)}$
($a^\dagger_{R(L)}$). In dipole and rotating wave approximation,
the interaction of the atom with the cavity field has the form
$H_{\rm AC}$, where $\sigma^+_1=\op{e_1}{g_1}$,
$\sigma^+_2=\op{e_2}{g_2}$ and
$\sigma^-_j=(\sigma^+_j)^{\dagger}$, $j=1,2$. Furthermore, $g$
denotes the single photon Rabi frequency, and $k$ is the wave
vector of the laser light. The free Hamiltonian for the external
driving field is denoted by $H_{\rm F}$ and $H_{\rm CF}$ describes
the interaction of this field with the two cavity modes, where
$\kappa$ is the photon loss rate through the cavity mirrors. The
bosonic annihilation (creation) operators for the field in the
mode $\omega$, $b_{\omega,R(L)}$ ($b_{\omega,R(L)}^\dagger$)
fulfill the commutation relations
$[b_{\omega,R(L)},b_{\omega',R(L)}^\dagger]=\delta(\omega-\omega')$.
Finally, the term $H_{\rm sys-bath}$ describes the interaction of
the atom with a photonic bath $H_{\rm bath}$, which leads to
spontaneous emission. These terms are of standard form and can be
found e.g.~in \cite{Walls,QN}.

We assume the electric field driving the two cavity modes to be
coherent, representing a classical field with amplitudes
$\beta_{\rm in, R(L)}(t)$. As described in detail in
\ref{Coherent} we go to a picture where the incident driving field
explicitly enters the Hamiltonian. This results in a vacuum input
state for the cavity and the additional driving term
\begin{equation}\label{HCF}
    \widetilde{H}_{\rm CF}=\rmi\sqrt{2\kappa}\left(\beta_{\rm in}(t)
    A_+^{\dagger}-A_+\beta_{\rm in}(t)^*\right)
\end{equation}
in the Hamiltonian. Here we have defined $A_\pm = \left(a_R\pm
a_L\right)/\sqrt{2}$, and furthermore assumed symmetric driving
$\beta_{{\rm in },R}(t)=\beta_{{\rm in },L}(t)\equiv\beta_{\rm
in}(t)$.

The dynamics of the system thus obeys a master equation (see
e.g.~\cite{Walls})
\begin{equation}\label{ME}
    \dot\rho=-\rmi\left[H_{{\rm
    sys}},\rho\right]+\mathcal{L}(\rho),
\end{equation}
where the Hamiltonian of the system is given by $H_{\rm
sys}=H_{\rm A} +H_{\rm C}+H_{\rm AC}+\widetilde{H}_{\rm CF}$.
The Liouville superoperator $\mathcal{L}$ has the form
\begin{eqnarray}\label{Liouville1l}
    \mathcal{L}(\rho)&=& \frac{\kappa}{2}\sum_{i=\pm}\left(2A_i\rho
    A^{\dagger}_{i}-A^{\dagger}_{i} A_i\rho-\rho A^{\dagger}_{i}
    A_i\right) \\ \label{Liouville2l}
    &&+\frac{\gamma}{2}\sum_{i=\pm}\left(2\sigma^-_i\rho\sigma^+_i-\sigma^+_i\sigma^-_i\rho-\rho\sigma^+_i\sigma^-_i\right).
\end{eqnarray}
The first line Eq.~(\ref{Liouville1l}) describes loss due to the imperfectness of the
cavity mirrors, and the second line Eq.~(\ref{Liouville2l}) is due
to the spontaneous emission of the atom with emission rate
$\gamma$. Note that based on the choice of atomic states and laser
and cavity polarizations, the laser and cavity couplings as well
as the spontaneous emission rates are assumed to be identical for
the two qubit states.

Next we decompose the cavity mode operators $A_\pm$ into a
$\mathbbm{C}$-number $\alpha_\pm=\langle A_\pm \rangle$ describing
the coherent part and an operator part $\dA_\pm=A-\alpha$
describing the fluctuations. The mean values
$\EV{A_\pm}=\alpha_\pm$ are calculated in the steady state. From
Eq.~(\ref{ME}) the respective equations of motion for $\EV{A_\pm}$
can be easily obtained, yielding
\begin{eqnarray}\label{EValpha}
    &\dt{\EV{A_+}}=(\rmi\Delta_c-\frac{\kappa}{2})\EV{A_+}+\sqrt{2\kappa}\beta_{\rm in}, \nonumber\\
    &\dt{\EV{A_-}}=(\rmi\Delta_c-\frac{\kappa}{2})\EV{A_-}.
\end{eqnarray}
From the steady state solution of Eqs.~(\ref{EValpha}) we find
\begin{equation}
    \alpha_+=\frac{\sqrt{2\kappa}\beta_{\rm in}}
    {-\rmi\Delta_c+\frac{\kappa}{2}}\equiv\alpha,\qquad
    \alpha_-=0.
\end{equation}
Applying this transformation to the system Hamiltonian we find
$H_{\rm sys}=H_{\rm A}+H_{\rm C}+\tilde H_{\rm int}$, where in
$H_{\rm C}$ the operators $A_\pm$ are replaced by $\dA_\pm$ and
$\tilde H_{\rm int}$ is given by
\begin{eqnarray}
   \tilde H_{\rm int}=\sqrt{2}g\left[\left((\delta
   A_++\alpha)\cos{kx}+\rmi
   \dA_-\sin{kx}\right)\left(\sigma^+_1+\sigma^+_2\right)\right.\nonumber\\
   \left.\qquad+{\rm h.c.}\right].\nonumber
\end{eqnarray}
The Liouville operator $\mathcal{L}$ reduces to
\begin{eqnarray}
    \mathcal{L}(\rho)&=& \frac{\kappa}{2}\sum_{i=\pm}\left(2\dA_i\rho
    \dA^{\dagger}_{i}-\dA^{\dagger}_{i} \dA_i\rho-\rho \dA^{\dagger}_{i}
    \dA_i\right),
\end{eqnarray}
where we have assumed large detunings $\Delta$ and consequently
neglected the spontanous emission, as it is suppressed by a factor
$\propto g^2/\Delta^2$. Furthermore, in this approximation we can
adiabatically eliminate the excited states $\ket{e_1}$ and
$\ket{e_2}$ and obtain
\begin{eqnarray}\label{Hefftrig}
    H_{{\rm eff}}&=\mathbbm{1}_i\otimes\left(\frac{p^2}{2m}-\Delta_c\left(\dA^{\dagger}_
    {+}\dA_++\dA^{\dagger}_{-}\dA_-\right)\right.\\
     &\left.+\frac{2g^2}{\Delta}\left(\alpha^2\cos^2kx+\rm
    i\alpha(\dA^{\dagger}_{-}-\dA_-)\cos{kx}\sin{kx}+\dA^{\dagger}_-\dA_-\sin^2kx\right)\right),\nonumber
\end{eqnarray}
where we have only kept the leading terms in $\dA$, assuming small
fluctuations $|\EV{\delta A}|\ll |\alpha|$. In
Eq.~(\ref{Hefftrig}) the operator $\mathbbm{1}_i=\op{g_1}{g_1}+
\op{g_2}{g_2}$ which we do not write explicitly in the following.
Thus, due to our choice of symmetric coupling and dissipation, the
information encoded in the qubit (i.e.~the amplitudes of the two
internal ground states) is not affected by the time evolution
Eq.~(\ref{ME}). Furthermore, the operators $\dA_+$ and
$\dA^{\dagger}_+$ do not couple to any other degrees of freedom of
the system and, therefore, we drop them in the following.

\begin{figure}[htp]
    \begin{center}
        \includegraphics{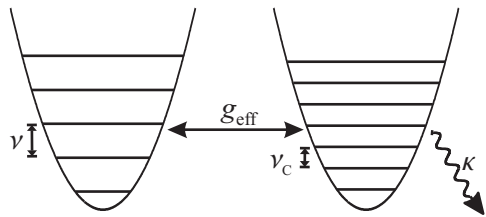}
        \caption{\label{fig:HO} The two coupled harmonic oscillators corresponding to the atomic motion and
        the cavity with frequencies $\nu$ and $\nu_c$, respectively. The coupling strength is
        denoted by $g_{\rm eff}$ and $\kappa$ is the photon loss rate through the cavity
        mirrors.}
    \end{center}
\end{figure}

The periodic optical lattice inside the cavity is given by the
term $V_0\cos^2{kx}$ in Eq.~(\ref{Hefftrig}) with a depth of
$V_0=2g^2\alpha^2/|\Delta|$. For a blue (red) detuned lattice and
a kinetic energy of the atom much smaller than the depth of the
optical lattice (i.e.~$\EV{p^2}/2m\ll V_0$), the atoms are trapped
near the (anti)nodes of the standing wave. Thus, we can
approximate each well of the lattice by a harmonic oscillator with
frequency $\nu=2 g \alpha k /\sqrt{m \Delta}$. In this
approximation $\mathcal{L}$ is given by
\begin{equation}\label{Liouville}
    \mathcal{L}(\rho)=\frac{\kappa}{2}\left(2a\rho
    a^{\dagger}-a^{\dagger}a\rho-\rho a^{\dagger}a\right),
\end{equation}
where we have introduced $a\equiv\dA_-$ and the system Hamiltonian
reduces to
\begin{equation}\label{Hfin}
    H_{\rm sys}=\nu   c^{\dagger}c+\nu_c a^{\dagger}a+\rmi g_{{\rm
    eff}}\left(c+c^{\dagger}\right)\left(a^{\dagger}-a\right).
\end{equation}
The bosonic creation (annihilation) operator $c^{\dagger}$ ($c$)
describes the quantized motion of the atom in a harmonic potential
of frequency $\nu$ coupled to the cavity mode $a$ with the
effective coupling strength $g_{{\rm eff}}= \nu/2\alpha\eta$, as
shown in Fig.~\ref{fig:HO}. In Eq.~(\ref{Hfin}) we also used the
definitions $\nu_c=-\Delta_c+\frac{2g^2}{\Delta}$ for the
frequency of the cavity mode $a$, and $\eta = \sqrt{k^2/2m\nu}$
the Lamb-Dicke parameter. The shift $2g^2/\Delta$ can be
interpreted as a refractive index due to the back-action of the
atom on the cavity field. In the limit $g/\nu\ll 1,\nu = \nu_c$ we
can apply the rotating wave approximation (RWA) to
Eq.~(\ref{Hfin}), and find
\begin{equation}\label{RWA}
    H_{\rm RWA}=\nu c^{\dagger}c+\nu_c a^{\dagger}a+\rmi g_{{\rm
    eff}}\left(ca^{\dagger}-ac^{\dagger}\right).
\end{equation}

We finally turn to the physical interpretation of our results. The
coupled oscillator Hamiltonian Eq.~(\ref{Hfin}) describes the
coherent transfer of phonons (vibrational excitations) to cavity
photons. Due to its high frequency, the optical cavity mode is
coupled to an effective zero temperature heat bath. Cavity phonons
will decay through the mirror according to Eq.~(\ref{Liouville}),
dissipating the energy.

A physically equivalent picture can be given in terms of a {\em
dissipative optical lattice} according to Eq.~(\ref{Hefftrig}). In
the ring cavity we have two counterpropagating modes, which we
rewrite as $\cos kx$ (+) and $\sin kx$ (-) standing wave modes. By
driving the cavity symmetrically from the outside with two lasers,
the standing wave $\cos kx$ will have a large occupation of
phonons $|\alpha|^2$, giving rise to an optical lattice potential,
while the not-driven $\sin kx$ mode will be in a vacuum state.
Motion of an atom in the optical lattice will redistribute the
photons among the two modes. In particular, atomic motion will
transfer photons to the (initially empty) $\sin kx$ mode, i.e.~the
optical lattice potential will be slightly shifted by atomic
motion.  On the other hand, the photons occupying the $\sin kx$
mode can decay through the cavity mirror, giving rise to a
damping. Thus we have the picture, where the atomic motion in the
optical wells acts back on the lattice to produce a (small
amplitude) displacement of the lattice. These oscillations of the
optical lattice will damp to their equilibrium position by the
cavity decay, thus cooling the motion of the atom. With an
appropriate choice of qubit states we assure that the dissipative
optical lattice potential seen by the two qubit states are
identical. This ensures that the cooling does not destroy the
coherence of the qubit.

Finally, we note that our laser cooling model maps on to a
remarkably simple mathematical model of coupled harmonic
oscillators with linear damping. Thus we find a straightforward
solution in terms of Gaussian states of the two modes. Below we
will extend the model to $N$ qubits, and then to the case of a
time dependent moving lattice to illustrate the cooling dynamics.

\subsubsection{N Qubits}\label{Nqubits}

The derivation of the $N$-particle Hamiltonian is similar to the
calculations presented in Sec.~\ref{singleQ} and leads to
\begin{equation}\label{H_Nfin}
    H=\nu\sum_{\mu=1}^Nc_{\mu}^{\dagger}c_{\mu}+\nu_c(N) a^{\dagger}a+\rmi
    g_{\rm eff}\sum_{\mu=1}^N\left(c_{\mu}+c_{\mu}^{\dagger}\right)\left(a^{\dagger}-a
    \right),
\end{equation}
where the index $\mu$ labels the particles. All parameters apart
from $\nu_c(N)=-\Delta_c+2g^2N/\Delta$ and the refractive index
now given by $2g^2N/\Delta$ remain unchanged. In
Eq.~(\ref{H_Nfin}) we have assumed the effective coupling constant
to be the same for each particle ($g_{\rm eff,\mu} = g_{\rm
eff}$). The Hamiltonian Eq.~(\ref{H_Nfin}) describes N harmonic
oscillators corresponding to the quantized motion of the atoms
coupled to the single cavity mode $a$. The coupling term in
Eq.~(\ref{H_Nfin}) has exactly the same form as in the single
qubit case Eq.~(\ref{Hfin}) if we define the center of mass (COM)
annihilation (creation) operator $C$ ($C^\dagger$) given by
$C=1/\sqrt{N} \sum_{\mu}c_{\mu}$ and introduce the coupling
constant $\tilde g = \sqrt{N}g_{\rm eff}$. The resulting form of
Eq.~(\ref{H_Nfin}) already indicates that only the collective mode
$C$ will be cooled in our approximative description. We note that
previous publications \cite{HorakN,CollCool} have already shown in
a semiclassical approximation, that also the (N-1) relative
vibrational modes can be cooled with our setup, but on a much
larger timescale.

As will be discussed in detail below, transport of $N$ atoms in an
accelerating lattice leads to coherent oscillations of the
center-of-mass mode of all the atoms. Our cooling scheme cools
this particular collective mode with high efficiency due to the
collective enhancement $g_{\rm eff} \rightarrow g_{\rm eff}
\sqrt{N}$ of the phonon-photon.

\subsection{Optical Lattice in a Standing Wave Cavity}

\begin{figure}[htp]
  \begin{center}
    \includegraphics{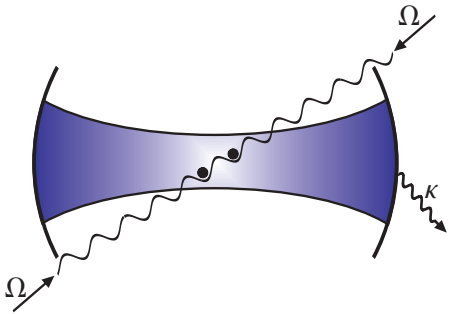}
    \caption{\label{lascav} Schematic experimental setup as described in the text.}
  \end{center}
\end{figure}

As a possibly more convenient setup, we replace the ring cavity of
Sec.~\ref{model_ring} by a standing wave cavity providing the
cooling. We assume the optical lattice holding to the atoms to be
produced by two counter-propagating laser beams, which we shine
into the resonator at a small angle, as shown in
Fig.~\ref{lascav}, whereas in the previous system the driving
field entered the cavity through the cavity mirrors.
The total Hamiltonian takes on the form $ H=H_{\rm A} +H_{\rm
C}+H_{\rm AC}+H_{\rm AL}$, where
\begin{eqnarray}
    &H_{\rm A} =\frac{p^2}{2m}+(-\Delta)(\op{e_1}{e_1}+\op{e_2}{e_2}),\\
    &H_{\rm C}=(-\Delta_c)a^{\dagger}a,\nonumber\\
    &H_{\rm AL}=\frac{\Omega}{2}\rmi\sin{kx}\left(\sigma^+-\sigma^-\right),\nonumber\\
    &H_{\rm
     AC}=g\cos{k_cx}\left(a\sigma^++a^{\dagger}\sigma^-\right),\nonumber
\end{eqnarray}
and $\sigma^\pm = (\sigma_1^\pm+\sigma_2^\pm)$.
The Hamiltonians $H_{\rm A}$ and $H_{\rm C}$ denote the energy of
the atom and the cavity with bosonic creation and annihilation
operators $a^{\dagger}$ and $a$, respectively. The interaction of
the laser with Rabi frequency $\Omega$ and the atom is denoted by
$H_{\rm AL}$ and $H_{\rm AC}$ describes the atom-cavity coupling
($g$ again denotes the single photon Rabi frequency).

Applying the same approximations as in Sec.~\ref{model_ring} we
find an identical model with the replacements $\nu \rightarrow
\tilde \nu= \Omega^2\eta^2/\Delta$ and $\nu_c \rightarrow \tilde
\nu_c= -\Delta_c+g^2/\Delta$, as well as the effective coupling
$g_{{\rm eff}} \rightarrow \tilde g_{{\rm eff}}=g\Omega\eta
/2\Delta$ in Eq.~(\ref{ME}). The Lamb-Dicke parameter $\eta$ is
$\eta = \sqrt{k/2m\nu}\sim\sqrt{k_c/2m\nu}$, where $k_c$ and $k$
denote the wave vectors of the cavity and the laser light,
respectively. A straightforward extension to $N$ qubits results
again in a Hamiltonian of the form (\ref{H_Nfin}).

The physical interpretation of this setup is very similar to the
previous one with the only difference that the optical lattice is
formed by two additional lasers instead of the cavity itself. In
this case the motion of the atoms transfers photons from the laser
field into the (initially empty) cavity. As in the previous setup
the decay of the cavity photons through the cavity mirrors leads
to cooling of the atomic motion without affecting the qubit state.

\subsubsection{Accelerated optical lattice}

Finally, we generalize the model to an accelerated optical
lattice. In the setup of  Fig.~\ref{lascav} the lattice can easily
be moved by introducing a time dependent relative phase $\phi(t)$
between the two counter-propagating laser beams
\cite{ColdColl,BlochMove}. We find a system Hamiltonian
\begin{eqnarray}
    H&=\nu c^{\dagger}c+\nu_c a^{\dagger}a-\frac{\nu\alpha}{2}
    \left(c+c^{\dagger}\right)-\rmi g\alpha
    \cos{(\phi(t))}\left(a^{\dagger}-a\right)\nonumber\\
    &+\rmi g\cos{(\phi(t))}\left(c+c^{\dagger}\right)\left(a^{\dagger}-a\right)
    +\frac{\nu^2}{4}\alpha^2,\label{HnonRWAacc}
\end{eqnarray}
with $\nu_c=-\Delta_c+g^2 \cos^2{(\phi)}/\Delta$,
$\alpha=\phi/\eta$. The values for $\nu$, $g$ and $\eta$ are
defined as above.

\section{Results}\label{Cool}

Below we solve the cooling equations Eq.~(\ref{ME}). In
Sec.~\ref{analytics} we derive analytical expressions for the
atomic steady state temperature $T_f$ and the cooling time $\tau$.
In Sec.~\ref{numerics} we show examples for the numerical solution
of Eq.~(\ref{ME}) in the case of a static lattice as well as for
an accelerated lattice.

\subsection{Analytical Calculations}\label{analytics}

The master equation (\ref{ME}) consists of a quadratic system
Hamiltonian and a linear damping term. Thus, the Heisenberg
equations for the expectation values of the second order moments
can be found easily (see \ref{App_TE}, Eq.~(\ref{DEnonRWA})). From
these we calculate the atomic steady state temperature $T_F$
\begin{eqnarray}\label{Tfin}
    &k_BT_f \equiv E_0=\nu\EV{c^{\dagger}c}_{\rm SS}=\\
    &=\frac{g_{\rm eff}^2}{2\nu}+\frac{\kappa^2+4(\nu-\nu_c)^2}{16\nu_c}+
    \frac{8g_{\rm eff}^4\nu_c}{\nu\left[\kappa^2\nu+4\nu_c(-4g_{\rm
    eff}^2+\nu\nu_c)\right]},\nonumber
\end{eqnarray}
with $k_B$ the Boltzmann constant and $E_0$ the steady state
energy. We note that the system is always cooled to the ground
state in the limit $g_{\rm eff}\ll\nu\sim\nu_c$ (i.e.~where the
RWA Eq.~(\ref{RWA}) is valid).

The atomic energy $E_{\rm atom}(t)=\nu\EV{c^{\dagger}c}_t$ can be
written in the form
\begin{equation}\label{Eatom}
    E_{\rm atom}(t)=\sum_k c_k \rm e^{\lambda_kt}+E_0,
\end{equation}
where the $\lambda_k$ are eigenvalues (all having negative real
parts) obtained from the equations of motion and the coefficients
$c_k$ are found from the initial motional state of the atom. If
one term dominates in the sum Eq.~(\ref{Eatom}) then $1/|{\rm
Re}\lambda_k|$ with the corresponding eigenvalue $\lambda_k$ gives
the cooling time $\tau$. In general, however, the cooling time
will be determined by the eigenvalue $\lambda_k$ with the largest
real part unless the corresponding coefficient $c_k$ vanishes. We
therefore find an upper bound for the cooling time $\tau$ by
\begin{equation}\label{tau}
    \tau=\frac{1}{{\rm min}\{|{\rm Re}\lambda_k|\}}_k.
\end{equation}

In RWA we can easily calculate the eigenvalues $\lambda_k$ from
the equations of motion given in \ref{App_TE}, Eq.~(\ref{DERWA})
and find simple analytic estimate for the cooling time $\tau$ in
two interesting limits. In the "Doppler limit", i.e.~$g_{\rm
eff}\ll\nu\ll\kappa$, $\nu_c=\frac{\kappa}{2}$ we find
$\tau\approx \kappa/4g_{\rm eff}^2$ and
$k_BT_f\approx\frac{\kappa}{4}$, in close analogy to free space
Doppler cooling (see e.g.~\cite{Metcalf}). In the "sideband
limit", i.e.~$g_{\rm eff}\sim\kappa\ll\nu\sim\nu_c$, we obtain
$\tau\approx\frac{2}{\kappa}$ and $k_BT_f\approx g_{\rm
eff}^2/2\nu$. In the numerical examples in Sec.~\ref{numerics} one
can see that those analytical expressions are in very good
agreement with the exact numerical calculations. These results for
the single qubit case can be directly applied to the COM motion of
$N$ qubits by replacing $g\rightarrow\tilde g$.

\subsection{Numerical Results}\label{numerics}
In this section we show the numerical solution of Eq.~(\ref{ME})
for different parameter regimes in the case of a static lattice as
well as for an accelerated lattice.
\begin{figure}[htp]
    \begin{center}
        \includegraphics{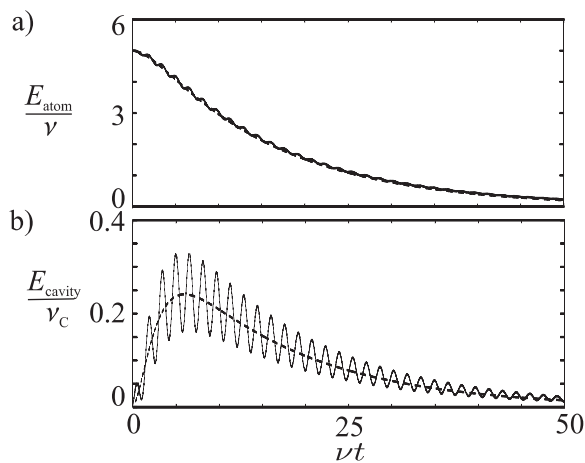}
        \caption{\label{fig:compare} a) Energy of the atomic motion $E_{\rm atom}$ and b) energy
        of the cavity mode $E_{\rm cavity}$ as a function of time $\nu t$. The parameters are
        chosen $\kappa=1$, $\nu=\nu_c=2$ and $g_{\rm eff}=1/8$. The dashed lines show the time evolution with
        the effective Hamiltonian $H_{\rm eff}$ in RWA, the solid lines are without the RWA.}
    \end{center}
\end{figure}
Fig.~\ref{fig:compare} shows the time evolution of the atomic
energy $E_{\rm atom}(t)$ and the energy of the cavity mode $a$,
$E_{\rm cavity} = \nu_c\EV{a^{\dagger}a}_t$, as a function of time
$\nu t$. Initially the system is assumed to be in a product state
with the cavity mode $a$ in its ground state and the mode for the
atomic motion in a coherent state. The time evolution of the two
energies shows that energy is transferred from the mode $c$ for
the atomic motion into the cavity mode. The energy of the cavity
mode decreases due to the photon loss through the cavity mirrors,
which leads to a cooling of the atomic motion as well. Furthermore
Fig.~\ref{fig:compare} shows a comparison between the time
evolution in RWA and without RWA. We can see, that in the RWA no
oscillations of the energies occur, i.e.~in this approximation the
energy is transferred smoothly from one mode into the other and
dissipates at a rate $\kappa$.
\begin{figure}[htp]
    \begin{center}
        \includegraphics{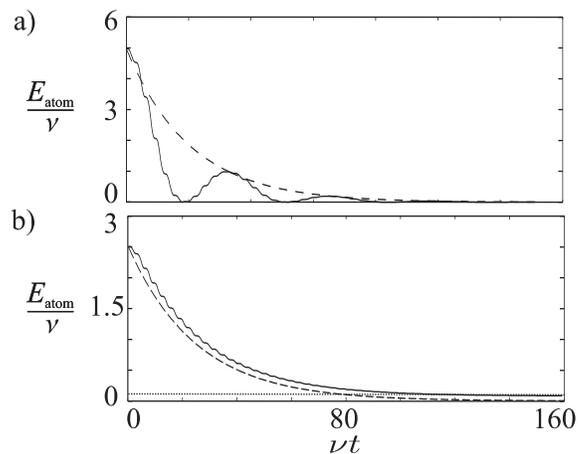}
        \caption{\label{fig:fit} Energy of the atomic motion $E_{\rm atom}/\nu$
        versus time $\nu t$. The parameters are chosen a) $\kappa=0.1$,
        $\nu=\nu_c=1$, $g_{\rm eff}=0.1$ and b) $\kappa=1$, $\nu=\nu_c=1$
        and $g_{\rm eff}=0.1$. The dashed curves show an exponential decay with
        using the analytical expression for $\tau$. The dotted line in b) gives
        the analytical value for the final energy $E_0/\nu$.}
    \end{center}
\end{figure}

In Fig.~\ref{fig:fit} we compare the analytical results for the
cooling rates and the final temperatures with the exact numerical
solution in two typical examples. One can see that the cooling
times which we have estimated by using Eq.~(\ref{RWA}), fit the
exact numerical calculations very well, and the final state
temperature (which is almost zero in Fig.~\ref{fig:fit}a coincides
exactly with the analytical value calculated from
Eq.~(\ref{Tfin}).

\begin{figure}[htp]
    \begin{center}
        \includegraphics{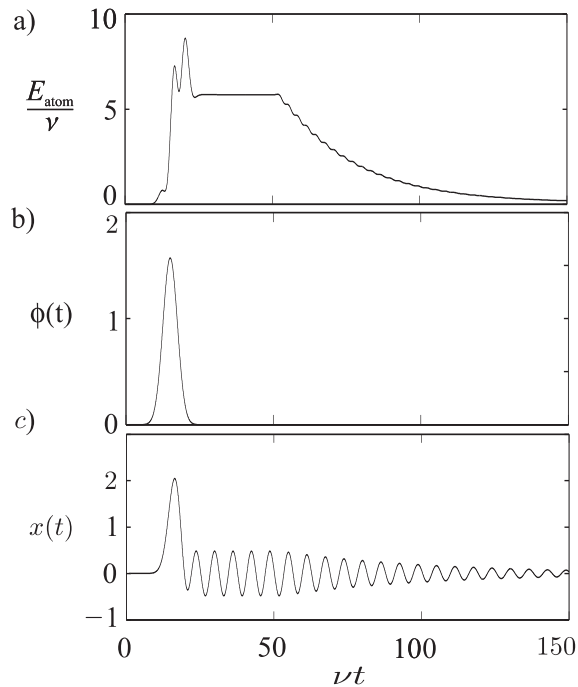}
        \caption{\label{fig:acc} a) Energy of the atomic motion $E_{\rm atom}/\nu$ versus time $\nu t$. Initially the
          atom is in its motional ground state, then motional excitations are created by non adiabatic acceleration of
          the lattice. Afterwards cooling is switched on at $\nu t=51$ with parameters $\kappa=1$, $\nu=\nu_c=1$ and
          $g_{\rm eff}=0.1$. b) Expectation value $x(t)=\EV{\hat x(t)}$ of the atomic position in
          dimensionless units and c) the relative phase $\phi(t)$ between the two lasers.}
    \end{center}
\end{figure}

Finally, Fig.~\ref{fig:acc} shows the excitation of the motional
state of an atom, initially assumed to be in its motional ground
state, by non adiabatic acceleration of the lattice. The chosen
velocity profile is typical for a two qubit quantum gate where one
atom is moved close to a second one for performing the gate
operation. While here we assume that the motional state is excited
during one such movement it might typically only be necessary to
re-cool the atom after a few gate operations have been performed.
After the motion has finished we turn on the cooling to bring the
atom back into its motional ground state. The results are
summarized in Fig.~\ref{fig:acc}. The time evolution of the atomic
energy
\begin{eqnarray}
   E_A = \nu\EV{c^{\dagger}c}-\frac{\nu\alpha}{2}\left(\EV{c}+\EV{c^{\dagger}}\right)
   +\frac{\nu^2}{4}\alpha^2-\frac{\rmi\nu\dot\alpha}{2}\left(\EV{c^\dagger}-
    \EV{c}\right)+\frac{\nu^2\dot\alpha}{4},
\end{eqnarray}
which includes the kinetic energy part due to the motion as well
as the vibrational excitation inside the trap is shown in
Fig.~\ref{fig:acc}a. The motion of the optical lattice is caused
by a time dependent relative phase $\phi(t)=kx_0(t)$
(c.f.~Fig.~\ref{fig:acc}b) with $x_0(t)$ the mean equilibrium
position of the atom. During the acceleration, the mean atomic
position $x(t)=\EV{\hat x(t)}$, shown in Fig.\ref{fig:acc}c,
follows its equilibrium position $x_0(t)$ quite well. After the
acceleration has stopped and the cooling has been turned on the
motion of the atom is cooled down to the temperature $T_f$
(Eq.\ref{Tfin}) within a few trap periods.

\section{Imperfections}\label{imperfect}
\begin{figure}[htp]
\begin{center}
    \includegraphics{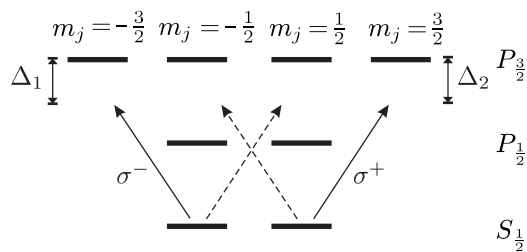}
    \caption{\label{fig:Alkali} Fine structure and laser configuration as
     described in the text.}
\end{center}
\end{figure}

In this section we discuss sources of imperfections leading to
qubit errors for a specific internal level structure of the
neutral atoms storing the qubits. We consider the fine structure
levels of an alkali atom as e.g.~Na or Rb shown in
Fig.~\ref{fig:Alkali} and the laser light ideally travelling along
the quantization axis $z$. We assume the qubit state $\ket{0}$
($\ket{1}$) to be encoded in the $m_j=-1/2$ ($m_j=+1/2$) state of
the $S_{\frac{1}{2}}$ manifold. This qubit state is coupled to the
excited state $m_j=-3/2$ ($m_j=+3/2$) of the $P_{\frac{3}{2}}$
manifold by a $\sigma^-$ ($\sigma^+$) polarized laser with
detuning $\Delta_1$ ($\Delta_2$). We consider two kinds of
imperfections (i) qubit flip errors due to misalignment of the
laser leading to small errors in the polarization vector
$\delta{\bf \epsilon}$ and (ii) stray magnetic fields $\delta {\bf
B}$ parallel to the quantization axis or equivalently imbalances
in the two Rabi frequencies leading to qubit phase errors,
i.e.~the accumulation of a relative phase between $\ket{0}$ and
$\ket{1}$ during the cooling process.

The probability $P_{\rm flip}$ for a spin flip error
($\ket{0}\leftrightarrow\ket{1}$) to occur in a given time
interval $\tau$ is found to be
\begin{equation}
    P_{\rm flip}=\frac{\Omega^4}{18\Delta^2}\tau^2
    \delta\epsilon_3^2,
\end{equation}
where $\delta\epsilon_3$ is the component of the polarization
vector in the direction of the quantization axis. Inserting
typical values $\tau\sim 20/\nu$ (cooling time found for the
parameters used in Fig.~\ref{fig:compare}) and $\delta\epsilon_3
\sim 10^{-4}$ we find $P_{\rm flip}\sim 10^{-3}$.

A magnetic field in $z$-direction $\delta B_3$ leads to a Zeeman
shift $\delta E=gm_j\mu_B\delta B$, with $\mu_B$ denoting the Bohr
magneton and $g$ the Land\'e factor of the different states. The
resulting difference in the transition frequencies from $\ket{0}
\leftrightarrow \ket{e_1}$ and $\ket{1} \leftrightarrow \ket{e_2}$
leads to two different Stark shifts $\Omega^2/\Delta_1$ and
$\Omega^2/\Delta_2$. Note that imbalances in the Rabi frequencies
or differences in the trapping potentials of the two qubit states
also lead to differences in these Stark shifts and can thus be
treated in an identical way. The relative phase accumulated in
time $\tau$ is given by
\begin{equation}
    \phi_{\rm rel}=\frac{\Omega^2}{2\Delta_2^2}\tau\mu_B\delta
    B_3.
\end{equation}
For the typical values used above and $\delta B_3 \sim 1G$ we
obtain a relative phase $\phi_{\rm rel}\sim 5 \times 10^{-3}$.

\section{Conclusion}

In this paper we have studied the nondestructive cooling of atomic
qubits in an optical lattice by coupling to a dissipative optical
cavity. In particular, we have proposed two specific experimental
configurations, where the optical lattice was produced in the
interior of an optical cavity, and the required dissipative
mechanism was provided by photon loss through the mirrors of the
cavity. In the first setup two counter-propagating laser beams
were driving two modes of a ring cavity, resulting in an
intra-cavity standing wave. We interpreted the resulting cooling
in terms of a dissipative optical lattice which by an appropriate
choice of atomic states became independent of the qubit state. In
the second setup the ring cavity was replaced by a standing wave
cavity and an optical potential was generated by two
counter-propagating external laser beams intersecting the cavity
at a small angle. In both cases the physics of cooling could be
described as two coupled oscillators consisting of the phonon mode
coupled parametrically to the optical cavity mode, which was
damped by cavity loss providing the dissipative element for the
cooling process.

Transport of $N$ atoms in the lattice gives rise to coherent
oscillations of the center-of-mass mode of all the atoms. Our
cooling scheme cools this particular collective mode with high
efficiency due to the collective enhancement $g_{\rm eff}
\rightarrow g_{\rm eff} \sqrt{N} $ of the coupling of the phonon
to the photon modes. We find cooling times which are similar to
those observed for other schemes \cite{Andrew} and are much
smaller than the relevant decoherence times for the internal
atomic states used in quantum information processing with neutral
atoms. Therefore our scheme can be applied repeatedly to cool the
motion of atoms which was excited during gate operations that
involved moving the atoms. While the present work has emphasized
application of this cooling scheme in a quantum information
context, coupling the motion of atoms in optical lattices to
photon modes of a cavity might also be interesting from the
perspective of quantum degenerate gases in optical lattices.

\ack{The authors thank Helmut Ritsch for discussions and a reading
of the manuscript, A.G. thanks the University of Oxford for
hospitality. This work was supported in part by the Austrian
Science Foundation FWF, E.U. Networks and the Institute for
Quantum Information.}

\appendix{
\section{Transformation to a Classical Driving Field}\label{Coherent}

We describe the transformation which allows the incident field to
be represented by a classical driving field in the
Hamiltonian~(\ref{Hsum}), similar to the work by Mollow
\cite{Mollow}. The state vector $\ket{\psi(t)}$ giving the system
and the driving field satisfies the Schr\"odinger equation
\begin{equation}
    \rmi\frac{\rm d}{{\rm d}t}\ket{\psi(t)}=H\ket{\psi(t)},
\end{equation}
where $H$ is given by Eq.~(\ref{Hsum}). The initital state is
assumed to be a product state of the system and a coherent field
state
\begin{equation}
    \ket{\psi(0)}=\ket{\psi_{\rm sys}(0)}\otimes D(\{\beta_{\omega,R},
    \beta_{\omega,L}\})\ket{\rm vac},
\end{equation}
where $\ket{\rm vac}$ is the vacuum state and
\begin{equation}
    D(\{\beta_{\omega,R},\beta_{\omega,L}\}) = \exp\left(\sum_{j=R,L}\int{\rm d}
    \omega\left(\beta_{\omega,j}b_{\omega,j}^\dagger-\beta_{\omega,j}^*
    b_{\omega,j}\right)\right)
\end{equation}
is the coherent displacement operator (see e.g.~\cite{Mollow}). In
order to transform the coherent incident field to a classical
driving field we apply the unitary transformation
\begin{equation}
    \ket{\psi(t)}\rightarrow D^\dagger(\{\beta_{\omega,R}\rme^{-\rmi\omega
    t},\beta_{\omega,L} \rme^{-\rmi\omega t}\})\ket{\psi(t)},
\end{equation}
which results in an initial vacuum state for the driving field and
in the transformation $b_{\omega,R(L)} \rightarrow
b_{\omega,R(L)}+\beta_{\omega,R(L)}\rme^{-\rmi\omega t}$. Assuming
symmetric driving
$\beta_{\omega,R}=\beta_{\omega,L}\equiv\beta_{\omega}$ the
Hamiltonian now has the form Eq.~(\ref{Hsum}) with the additional
term $\widetilde{H}_{\rm CF}$ defined in Eq.~(\ref{HCF}) and we
have used
\begin{equation}
    \beta_{\rm in}(t) = \frac{1}{\sqrt{2\pi}}\int{\rm d}\omega\rme^{\rmi
    (\omega_L-\omega)t}\beta_\omega.
\end{equation}

\section{Damping Equations}\label{App_TE}

The Heisenberg equations for the expectation values of the second
order moments can be obtained from the master equation (\ref{ME})
with the system Hamiltonian Eq.~(\ref{Hfin}), yielding
\begin{eqnarray}\label{DEnonRWA}
    \dt{\EV{c^{\dagger}c}}=g_{\rm eff}\left(\EV{ca}-\EV{ca^{\dagger}}-\EV{c^{\dagger}a}+\EV{c^{\dagger}a^{\dagger}}
    \right),\\
    \dt{\EV{a^{\dagger}a}}=g_{\rm eff}\left(\EV{ca}+\EV{ca^{\dagger}}+\EV{c^{\dagger}a}+\EV{c^{\dagger}a^{\dagger}}
    \right),\nonumber\\
    \dt{\EV{ca}}=g_{\rm eff}\left(\EV{c^{\dagger}c}+\EV{a^{\dagger}a}+\EV{c^2}-\EV{a^2}\right)+(-\rmi\nu-\rmi\nu_c-
    \frac{\kappa}{2})\EV{ca}+g_{\rm eff},\nonumber\\
    \dt{\EV{ca^{\dagger}}}=g_{\rm eff}\left(\EV{c^{\dagger}c}-\EV{a^{\dagger}a}+\EV{c^2}+\EV{a^{\dagger^2}}\right)+
    (-\rmi\nu+\rmi\nu_c-\frac{\kappa}{2})\EV{ca^{\dagger}},\nonumber\\
    \dt{\EV{c^{\dagger}a}}=g_{\rm eff}\left(\EV{c^{\dagger}c}-\EV{a^{\dagger}a}+\EV{c^{\dagger^2}}-\EV{a^2}\right)+
    (\rmi\nu-\rmi\nu_c-\frac{\kappa}{2})\EV{c^{\dagger}a},\nonumber\\
    \dt{\EV{c^{\dagger}a^{\dagger}}}=g_{\rm eff}\left(\EV{c^{\dagger}c}+\EV{a^{\dagger}a}+\EV{c^{\dagger^2}}-
    \EV{a^{\dagger^2}}\right)+(\rmi\nu+\rmi\nu_c-\frac{\kappa}{2})\EV{c^{\dagger}a^{\dagger}}+g_{\rm eff},\nonumber\\
    \dt{\EV{c^2}}=2 g_{\rm eff}\left(-\EV{ca}+\EV{ca^{\dagger}}\right)+2\rmi\nu\EV{c^2},\nonumber\\
    \dt{\EV{a^2}}=2 g_{\rm eff}\left(\EV{ca}+\EV{c^{\dagger}a}\right)+\left(-2\rmi\nu_c-\kappa\right)\EV{a^2},\nonumber\\
    \dt{\EV{c^{\dagger^2}}}=2 g_{\rm eff}\left(\EV{c^{\dagger}a}-\EV{ca^{\dagger}}\right)+2\rmi\nu\EV{c^{\dagger^2}},\nonumber\\
    \dt{\EV{a^{\dagger^2}}}=2 g_{\rm eff}\left(\EV{c^{\dagger}a^{\dagger}}+\EV{ca^{\dagger}}\right)+\left(
    2\rmi\nu_c-\kappa\right)\EV{a^{\dagger^2}}.\nonumber
\end{eqnarray}

In RWA the time evolution of the expectation values of the
quadratic operators is given by a homogeneous set of linear
differential equations, which can be written in the form
\begin{equation} \label{DERWA}
    \dt{\vec y(t)}=\mathcal{M}\vec y(t).
\end{equation}
Here $\vec y(t)$ is given by
\begin{equation}
    \vec y(t) = \left(\EV{c^{\dagger}c},\EV{a^{\dagger}a},\EV{ca^{\dagger}},
    \EV{c^{\dagger}a}\right)^{\rm T},
\end{equation}
and the time evolution matrix $\mathcal{M}$ has the form
\begin{eqnarray}
    \mathcal{M}=\left(
    \begin{array}{cccc}
        0& 0& -g_{\rm eff}& -g_{\rm eff}\\
        0& -\kappa& g_{\rm eff}& g_{\rm eff}\\
        g_{\rm eff}& -g_{\rm eff}& \rmi(\nu_c-\nu)-\frac{\kappa}{2}& 0\\
        g_{\rm eff}& -g_{\rm eff}& 0& \rmi(\nu-\nu_c)-\frac{\kappa}{2}\\
    \end{array}\right). \nonumber
\end{eqnarray}
}

\end{document}